\documentclass[fleqn,twoside]{article}
\usepackage{espcrc2}

\usepackage{graphicx}
\usepackage[figuresright]{rotating}

\graphicspath{{./Figures/}}                           

\newcommand{\etal}{\textit{\mbox{et al.\ }}}          
\newcommand{\vs}{\textit{\mbox{vs.\ }}}               

\renewcommand{\Re}{\mbox{Re}\,}                       
\newcommand{\Tr}{\mbox{Tr}}                           
\newcommand{\bc}{\texttt{bc}}                         
\newcommand{\fc}{\texttt{fc}}                         
\newcommand{\Fig}[1]{Fig.~\ref{#1}}

\newcommand{\Eq}[1]{Eq.~(\ref{#1})}

\newcommand{\beq}{\begin{equation}}
\newcommand{\eeq}{\end{equation}}
\newcommand{\bea}{\begin{eqnarray}}
\newcommand{\eea}{\end{eqnarray}}
\newcommand{\beas}{\begin{eqnarray*}}
\newcommand{\eeas}{\end{eqnarray*}}

\newcommand{\preprint}{\newline%
  \begin{picture}(0,0)
  \put(315,65){\rm\small HU--EP--05/75, LU-ITP 2005/024}
  \end{picture}}

\newcommand{\AmS}{{\protect\the\textfont2
  A\kern-.1667em\lower.5ex\hbox{M}\kern-.125emS}}

\title{
Landau gauge ghost and gluon propagators 
and the Faddeev-Popov operator spectrum 
\preprint
}
\author{
A. Sternbeck 
\address[HUB]{Humboldt-Universit\"at zu Berlin,
Institut f\"ur Physik, D-12489 Berlin, Germany}%
\thanks{Supported by the DFG-funded graduate school GK 271.},
E.-M. Ilgenfritz \addressmark[HUB]%
\thanks{Supported by DFG under FOR 465 / Mu 932/2.},
M. M\"uller-Preussker\addressmark[HUB]%
\thanks{Speaker at the Workshop on Computational Hadron Physics, 
Cyprus, September 2005.},
and
A. Schiller\address{Universit\"at Leipzig, Institut f\"ur Theoretische 
Physik, D-04109 Leipzig, Germany}
}      
\begin{document}

\begin{abstract}
In this talk we report on a recent lattice investigation of the
Landau gauge gluon and ghost propagators in pure $SU(3)$ lattice gauge
theory with a special emphasis on the Gribov copy problem. In the (infrared) 
region of momenta $~q^2 \le 0.3 ~\mathrm{GeV}^2~$ we find the corresponding 
MOM  scheme running coupling $~\alpha_s(q^2)~$ to rise in $q$. We also report 
on a first $SU(3)$ computation of the ghost-gluon vertex function showing 
that it deviates only weakly from being constant. In addition we study 
the spectrum of low-lying eigenvalues and eigenfunctions of the Faddeev-Popov 
operator as well as the spectral representation of the ghost propagator.
\vspace{1pc}
\end{abstract}

\maketitle

\section{INTRODUCTION}
The infrared behavior of the Landau gauge gluon and ghost 
propagators in QCD is closely related to the confinement 
phenomenon formulated in terms of the Gribov-Zwanziger horizon 
condition \cite{Zwanziger:1993dh_Gribov:1977wm} and the Kugo-Ojima
criterion \cite{Kugo:1979gm}. Assuming a constant ghost-gluon vertex
function the propagators also provide a nonperturbative determination of the 
running QCD coupling $~\alpha_s(q^2)~$ in a MOM scheme. They are a subject 
of intensive studies within the framework of truncated systems of
Dyson-Schwinger (DS) equations and  within the lattice approach.

In the infinite 4-volume limit the DS approach has led to an intertwined 
infrared power behavior of the gluon and ghost dressing functions 
\cite{vonSmekal:1997is_vonSmekal:1998_Alkofer:2000wg,Zwanziger:2001kw,%
Lerche:2002ep} 
\bea \nonumber
Z_{D}(q^2) &\equiv& q^2 D(q^2) \propto (q^2)^{2\kappa}, \\
Z_{G}(q^2) &\equiv& q^2 G(q^2) \propto (q^2)^{-\kappa}
\label{eq:infrared-behavior}
\eea
with the same $~\kappa\approx 0.59~$, i.e. a vanishing gluon propagator 
$D(q^2)$ in the infrared occurs in intimate connection with a diverging 
ghost propagator $G(q^2)$. As a consequence, the MOM scheme running coupling 
\beq
  \label{eq:runcoupling}
  \alpha_s(q^2) = \frac{g^2}{4\pi}~ Z_{D}(q^2)~ Z^2_{G}(q^2)
\eeq
is turning to a finite fix point in the infrared limit \cite{Lerche:2002ep}.
As we shall show below, our lattice simulations do not agree with this
expectation. $\alpha_s(q^2)$ seems to tend to zero in this limit. 

Possible solutions of the discrepancy have been discussed from different 
points of view in the recent literature.  
In \cite{Fischer:2002_2005} the DS approach has been studied on a finite 
4-torus with the same truncated set of equations as for the infinite volume. 
$~\alpha_s(q^2)~$ was shown to tend to zero for $~q^2 \to 0$ in one-to-one 
correspondence with what one finds on the lattice. This would indicate very 
strong finite-size effects and a slow convergence to the infinite-volume 
limit. However, we do not find any other indication for such 
a strong finite-size effect in our lattice simulations, except the convergence 
to the infinite-volume limit would be extremely slow. An alternative 
resolution of the problem has been proposed by Boucaud 
\etal \cite{Boucaud:2005ce}. These authors argue the ghost-gluon vertex 
in the infrared might contain $q^2$-dependent contributions which could modify 
the DS results for the mentioned propagators\footnote{We thank A. Lokhov 
for bringing the arguments in \cite{Boucaud:2005ce} to our attention.}.
However, recent detailed DS studies of the ghost-gluon vertex did not provide 
hints for such a modification \cite{Schleifenbaum:2004id_Alkofer:2004it}.
Thus, at present there is no solution of the puzzle. In any case more
thorough lattice investigations are desirable.

For $SU(2)$ extensive lattice investigations can be found in 
\cite{Bloch:2003sk}. Unfortunately, the authors did not reach the 
interesting infrared zone where the mentioned inconsistencies become visible.
The same holds for $SU(3)$ lattice computations of the ghost and gluon 
propagators as reported in \cite{Suman:1995zg,Furui:2003jr_Furui2004cx}.
We have been pursuing an analogous study for the $SU(3)$ case with the
special emphasis of the Gribov copy problem \cite{Sternbeck:2005tk}. 
Moreover, we have investigated the spectral properties of the Faddeev-Popov 
operator \cite{Sternbeck:2005vs}. The low-lying eigenmodes of the latter 
are expected to be intimately related to a diverging ghost propagator. 
In addition, in \cite{Sternbeck:2005qj} we have reported on a first
$SU(3)$ lattice computation of the ghost-ghost-gluon vertex at 
zero gluon momentum. We confirm, what has been found already for $SU(2)$, 
namely the data are quite consistent with a constant vertex function 
\cite{Cucchieri:2004sq}.

In this talk we will give a short review of these investigations. 
In the meantime larger lattice sizes and better statistics have been 
reached confirming our previous computations and conclusions.   

\section{GHOST AND GLUON DRESSING FUNCTIONS} 
We have simultaneously studied the gluon and ghost propagators in the
quenched approximation. $SU(3)$ gauge field configurations 
\mbox{$U=\{U_{x,\mu}\}$}
thermalized with the standard Wilson gauge action have been put into 
the Landau gauge by iteratively maximizing the gauge functional 
\beq
  F_{U}[g] = \frac{1}{4V}\sum_{x}\sum_{\mu=1}^{4}\Re\Tr
  \;{}^{g}U_{x,\mu}
  \label{eq:functional}
\eeq
with $g_x \in SU(3)$. It has numerous local maxima (Gribov copies), 
each satisfying the lattice Landau gauge condition $\partial_{\mu} A_{\mu}=0$. 
To explore to what extent this ambiguity has a significant influence on gauge
dependent observables, we have gauge-fixed each thermalized configuration
several times using the \emph{over-relaxation} algorithm, starting from
random gauge copies. For each configuration $U$, we have selected the
first (\fc{}) and the best (\bc{}) gauge copy (that with the largest
functional value) for subsequent measurements. For details we refer to
\cite{Sternbeck:2005tk}. 

It turns out that the Gribov ambiguity has no systematic influence on
the infrared behavior of the gluon propagator. This holds as long as
one restricts to periodic gauge transformations on the four-torus. 
In \cite{Burgio2005} the gauge orbits have been extended to the full 
gauge symmetry including also non-periodic ${\bf Z}(N)$ transformations. 
As a consequence the Gribov problem appears to be enhanced even for the 
gluon propagator. Here we restrict ourselves to the standard case of 
periodic gauge transformations. In marked contrast to the weak Gribov 
copy dependence of the gluon propagator the ghost propagator at lower 
momenta depends on the selection of gauge copies. (For a study of the 
influence of Gribov copies on the ghost propagator in the
$SU(2)$ case see \cite{Bakeev:2003rr}.)
\begin{figure}[ht]
 \includegraphics[width=7cm]{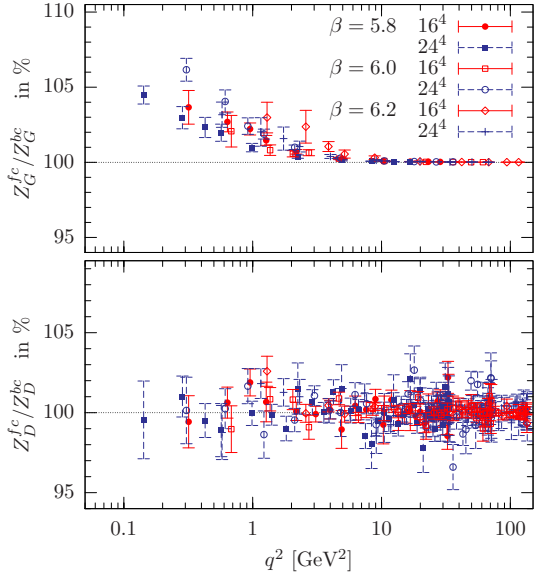} 
\vspace*{-0.5cm}
\caption{The ratios $Z^{\rm fc} /  Z^{\rm bc}$
  for the dressing functions $~Z_G~$ (upper panel) and $~Z_D~$ (lower panel)
  determined on first (\fc{}) and best (\bc{})
  gauge copies.}
\label{fig:ratio}
\end{figure}
The effect of the Gribov copies is illustrated in \Fig{fig:ratio}. There we 
have plotted the \fc{} - to - \bc{} ratios of the ghost and gluon dressing 
functions. Obviously for the ghost propagator the Gribov problem can cause 
$O(5 \%)$ deviations in the low-momentum region ($q<1$ GeV).
However, a closer inspection of the data for the ghost propagator indicates 
that the influence of Gribov copies becomes weaker for increasing lattice
size. This is in agreement with a recent claim by Zwanziger according to 
which in the infinite volume limit averaging over gauge copies in the
Gribov region should lead to the same result as over copies restricted to 
the fundamental modular region \cite{Zwanziger:2003cf}. In \cite{Burgio2005}
similar indications have been found for $SU(2)$.
\begin{figure}[ht]
 \includegraphics[width=7.0cm]{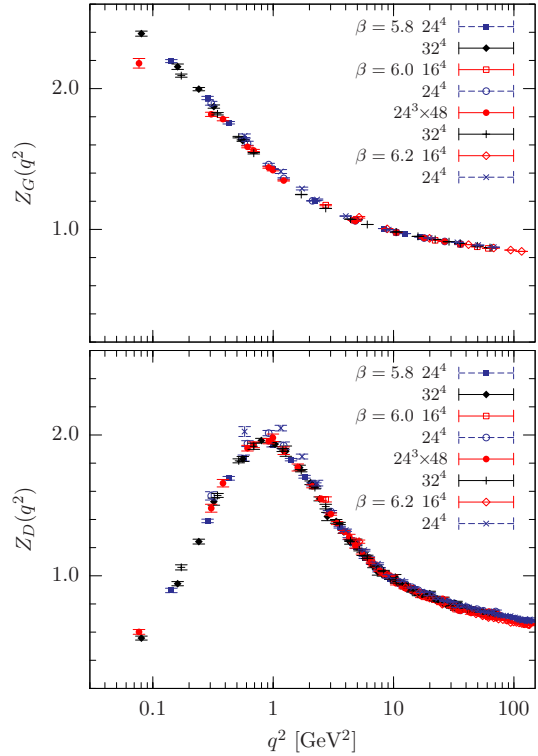} 
\vspace*{-0.5cm}
\caption{The ghost dressing function $~Z_G~$ (above) and 
  the gluon dressing function $Z_D$ (below) \vs momentum $q^2$ 
  both measured on \fc{}{} gauge copies.} 
\label{fig:ghost_gluon}
\end{figure}
\Fig{fig:ghost_gluon} shows the ghost and gluon dressing functions  
versus $q^2$. In contrast to our previous papers 
\cite{Sternbeck:2005tk,Sternbeck:2005qj} we have plotted only
\fc{}-copy results, where we have included some new data for lattices 
$24^3 \times 48$ and $32^4$ at $\beta=6.0$.
Both propagators have been renormalized separately for each $\beta$ 
with the normalization condition $Z=1$ at $q = \mu = 3$ GeV. 
Note the deviation of the data for the asymmetric lattice 
$24^3 \times 48$. In principle, for better gauge copies the curve 
can become only less singular. 
Still it is very difficult to extract a possible power behavior
at low $q^2$ in comparison with \Eq{eq:infrared-behavior}.

\section{THE GHOST-GLUON VERTEX AND THE RUNNING COUPLING}
\begin{figure}[ht]
  \includegraphics[height=5.5cm]{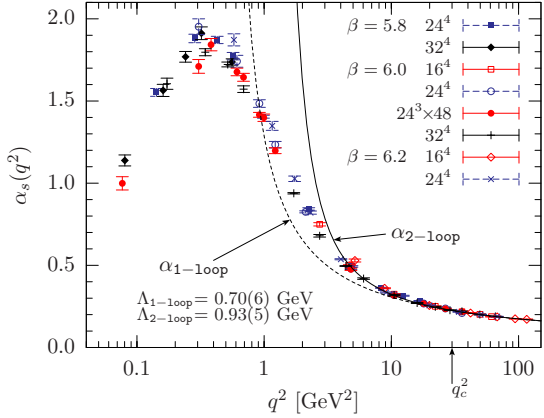} 
\vspace*{-0.5cm}
  \caption{The momentum dependence of the running coupling
    $\alpha_{s}(q^2)$ measured on \fc{} gauge copies.} 
\label{fig:alpha}
\end{figure}

In \Fig{fig:alpha} we present the combined result for the running coupling 
(\ref{eq:runcoupling}). Fits to the 1-loop and 2-loop running coupling 
are also shown.
\begin{figure}[ht]
  \includegraphics[height=5.5cm]{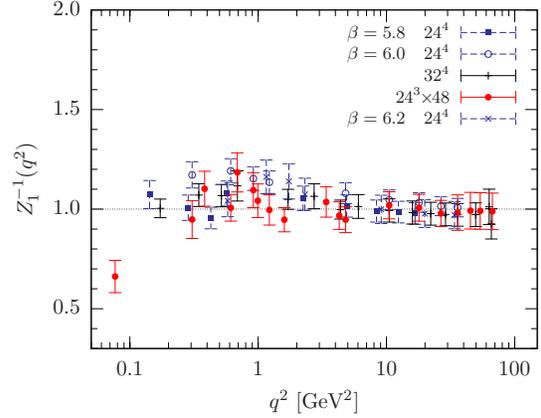}
\vspace*{-0.5cm}
  \caption{The inverse ghost-ghost-gluon vertex
    renormalization function $Z_1^{-1}(q^2)$ measured on \fc{} gauge 
    copies and normalized to unity at $q=3$ GeV.} 
\label{fig:vertex}
\end{figure}
Our new data points confirm our previous observation that the 
running coupling monotonously decreases with decreasing momentum in the 
range $q^2 < 0.4$~GeV$^2$. We have plotted \fc{} copy results, showing
that this behavior is not a matter of selecting better gauge copies. The nice
coincidence of the data for various lattice sizes (except the points 
refering to the asymmetric lattice) makes it hard to see 
how finite-volume effects can be blamed for the 
disagreement with the DS continuum result \Eq{eq:infrared-behavior}. 
Thus, the arguments as given in \cite{Boucaud:2005ce} should be checked 
in order hopefully to resolve the conflict.    
  
One can ask, whether the ghost-ghost-gluon vertex renormalization function 
$Z_1(q^2)$ is really constant at lower momenta. A recent investigation 
of this function defined at vanishing gluon momentum for the $SU(2)$ case 
\cite{Cucchieri:2004sq} supports that \mbox{$Z_1(q^2)\approx 1$} at least 
for momenta larger than 1~GeV. We have performed an analogous study for 
$Z_1(q^2)$ in the case of $SU(3)$ gluodynamics. Our (still preliminary) 
results are presented in \Fig{fig:vertex}. There is a slight variation
visible in the interval $0.3 ~\mathrm{GeV^2} \le q^2 \le 5 ~\mathrm{GeV^2}$.
But this weak deviation from being constant will not have a dramatic 
influence on the running coupling. 
 
\section{LOW-LYING MODES OF THE FADDEEV-POPOV OPERATOR}

\begin{figure}[ht]
  \includegraphics[width=7cm]{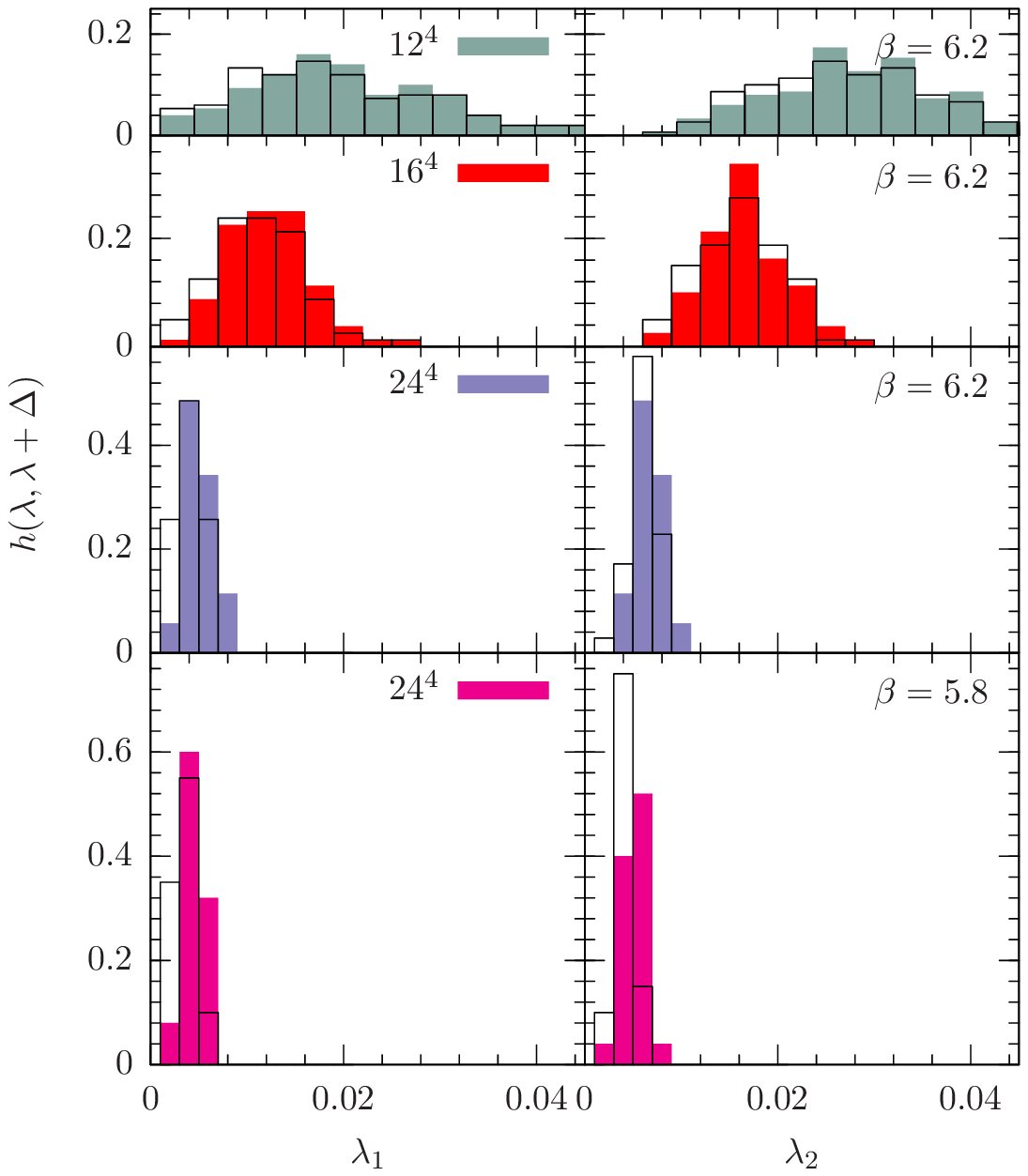}
\vspace*{-0.5cm}
  \caption{The distribution $h(\lambda)$ of the lowest
    (left panels) and second lowest (right panels) Faddeev-Popov 
    eigenvalues $\lambda_1$ and $\lambda_2$. Filled columns refer to 
    \bc{} gauge copies, while open columns refer to \fc{} copies.}
  \label{fig:fps_lowest}
\end{figure}
\begin{figure}[ht]
  \includegraphics[width=7cm]{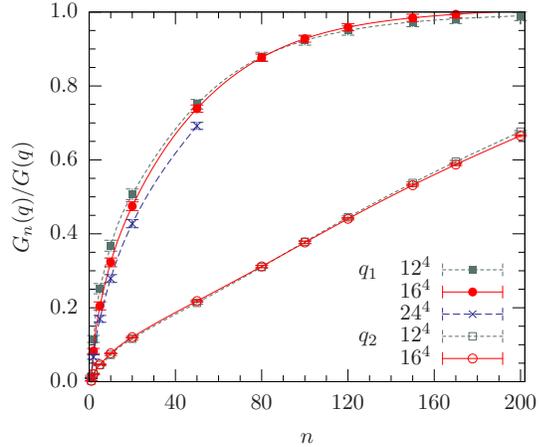}
\vspace*{-0.5cm}
  \caption{The ratio of the ghost propagator $G_n(q^2)$ approximated 
    in accordance with \Eq{eq:Def-ghost-q-by-spectrum}
    to the corresponding full estimator $G(q^2)$ as a function of $n$ 
    for the lowest $q_1$ and second lowest momentum $q_2~$ 
    ($\beta=6.2$, lattice sizes $12^4$, $16^4$ and $24^4$).}
  \label{fig:ghfps_n}
\end{figure}

Now let us discuss the spectrum of the low-lying eigenvalues
$\lambda_i$ of the Faddeev-Popov \mbox{(F-P)} operator. For each maximum of
the gauge functional $F_U[g]$, besides of its 
discarded eight trivial zero modes, the eigenvalues of the \mbox{F-P} operator 
are positive. The corresponding gauge field configuration is said to lie
within the Gribov region. If the lowest non-trivial eigenvalue tends to zero
the configuration approaches the so-called Gribov horizon.   

Having the lowest $n$ eigenvalues and eigenvectors determined,
one can construct an approximation to the ghost propagator 
\beq
  G_n(q^2) = \sum_{i=1}^{n}
  \frac{1}{\lambda_i}\,\vec{\Phi}_i(k)\cdot\vec{\Phi}_i(-k) 
\label{eq:Def-ghost-q-by-spectrum}
\eeq
with $\vec{\Phi}_i(k)$ being the Fourier transform of the $i$-th eigenmode
($k$ denotes the lattice momentum).
If all $n=8V$ eigenvalues and eigenvectors were known, 
the ghost propagator $~G(q^2)~$ would be completely determined.
In the recent literature~\cite{Greensite:2004ur,Gattnar:2004bf} it is
stated that the \mbox{F-P} operator acquires very small eigenvalues in the
presence of vortex excitations. In any case, according to a
popular belief it is an enhanced density of eigenvalues near zero 
which causes the ghost propagator to diverge stronger
than $1/q^2$ near zero momentum \cite{Zwanziger:2003cf}.

At $\beta=6.2$ we have extracted the 200 lowest (non-trival)
eigenvalues and their corresponding eigenfunctions for lattice sizes  
$12^4$, $16^4$ as well as 50 modes for $24^4$. Additionally, 
90 eigenvalues have been 
extracted on a $24^4$ lattice at $\beta=5.8$. This allows us to check
how the low-lying eigenvalues are shifted towards $\lambda=0$ as the
physical volume is increased.  In order to clarify how the choice of
gauge copy influences the spectrum, the eigenvalues and eigenvectors
have been extracted separately on our sets of \fc{} and \bc{}
gauge-fixed configurations.

In \Fig{fig:fps_lowest} the distributions of the
lowest $\lambda_1$ and second lowest $\lambda_2$ eigenvalues of the 
\mbox{F-P} operator are shown for different (physical) volumes.
There $h(\lambda,\lambda+\Delta)$ represents the
average number (per configuration) of eigenvalues 
found in the intervall $[\lambda,\lambda+\Delta]$. 
Open bars refer to the distribution on \fc{} gauge copies
while full bars to that on \bc{} copies.
From this figure it is quite obvious that both
eigenvalues, $\lambda_1$ and $\lambda_2$, are shifted to lower values as
the physical volume is enlarged. In conjunction the spread of
$\lambda$ values decreases. We have found that the center of those
distributions tends towards zero faster than $ 1/L^{2}$. Here $L$
refers to the linear lattice extension in physical units. For example,
we have found $\langle\lambda_{1}\rangle(L)\propto 1/L^{2+\varepsilon}$ with
$\varepsilon\approx0.16(4)$. It is also visible that 
the eigenvalues $\lambda_1$ and $\lambda_2$ on \fc{} gauge
copies are on average lower than those obtained on
\bc{} copies. 

\Eq{eq:Def-ghost-q-by-spectrum} suggests that low-lying eigenvalues 
and eigenvectors have a major impact on the ghost 
propagator at the lowest momenta. Actually, this is not so easy to predict.
We have studied, to what extent the lowest $n$ modes saturate the estimator 
of the ghost propagator $G_n(q^2)$ on a given set of gauge copies. 
We show the result in \Fig{fig:ghfps_n} for the lowest ($q_1$) and the second 
lowest momentum ($q_2$) available on lattices of different size at $\beta=6.2$. 
The estimates for $G_n(q^2)$ have been divided by the
full (conjugate gradient) estimator $G(q^2)$ in order to compare the 
saturation for different volumes. 
Considering first the lowest momentum $q_1$ we observe from
\Fig{fig:ghfps_n} that the rates of convergence differ,
albeit slightly, for the three different lattice sizes. The relative
deficit rises with the lattice volume. 
In order to reproduce the ghost propagator within a few percent, 
$150 \ldots 200$ eigenmodes turn out to be sufficient. 
For the second lowest momentum $q_2$ even 200 eigenmodes are far from
saturating the result.

\section{CONCLUSIONS}
We have studied the low-momentum region of Landau gauge
gluodynamics using Monte Carlo simulations with the Wilson plaquette 
action and various lattice sizes from $16^4$ to $32^4$. 
For the inverse bare coupling we haven chosen $\beta=5.8$, $6.0$ 
and $6.2$. In this way we have reached momenta down to 
$q^2 \simeq 0.1 \mathrm{GeV}^2$. We have presented data mainly for the
\fc{} gauge copies, but we have shown the ghost propagator to become
less singular within a $O(5\%)$ deviation, when better gauge copies are
taken. We have found indications that the Gribov effect weakens as the 
volume is increasing.
Towards $q \to 0$ in the infrared momentum region, the gluon dressing 
function was shown to decrease, while the ghost dressing function turned out
to rise. However, the connected power laws predicted by the infinite-volume 
DS approach cannot be confirmed from our data. Correspondingly, the behavior 
of the running coupling $\alpha_s(q^2)$ in a suitable momentum subtraction 
scheme (based on the ghost-ghost-gluon vertex) does not approach the 
expected finite limit.  Instead, this coupling has been found to decrease 
for lower momenta after passing a turnover at $q^2\approx0.4$~GeV$^2$. 

In addition, we have investigated the spectral properties of the \mbox{F-P}
operator and its relation to the ghost propagator for various lattice sizes. 
As argued in \cite{Zwanziger:2003cf} we have found the low-lying eigenvalues 
to be shifted towards zero (the configurations approaching the Gribov horizon) 
as the volume is increased.  The low-lying eigenvalues extracted on 
\bc{} gauge copies (those with the largest functional value) are larger 
on average than those on \fc{} copies. Thus, for finite volumes better 
gauge-fixing (in terms of the gauge functional) keeps the configurations 
at larger distance from the Gribov horizon. The study of the contributions 
to the ghost propagator coming from the lowest eigenmodes shows that even
at the lowest momentum one needs a large number of modes in order to saturate
the propagator. For increasing volumes and larger lattice momenta
the convergence of the eigenmode expansion becomes even slower.

\medskip
{\small
All simulations have been done on the IBM pSeries 690 at HLRN.  
We are grateful to Hinnerk St\"uben for providing parts of his 
code. One of us (M. M.-P.) acknowledges useful discussions with 
V.K. Mitrjushkin.}


\end{document}